\documentclass[10pt,aps,prl,groupedaddress,superscriptaddress,showpacs,
twocolumn
]{revtex4-1}

\usepackage{amssymb}
\usepackage{amsmath}
\usepackage{latexsym}
\usepackage{revsymb}
\usepackage{graphicx}

\def\Z{\hbox{{\sf Z}\kern-0.4em {\sf Z}}}

\begin{document}




\title{The Density of Surface States in Weyl Semimetals
}

\author{Alexander P. Protogenov}
 \affiliation{Institute of Applied Physics of the RAS, Nizhny Novgorod 603950,
Russia}
 \affiliation{ Donostia International Physics Center
(DIPC), 20018 San Sebasti\'an/Donostia, Spain}

\author{Valery A. Verbus}
 \affiliation{Institute for Microstructures of the RAS, Nizhny Novgorod 603950,
Russia}
\affiliation{National Research University Higher School of Economics,
Nizhny Novgorod 603155, Russia}
\author{Evgueni V. Chulkov}
\affiliation{Departamento de F\'isica de Materiales, Facultad de Ciencias Qu\'imicas,
Uviversidad del Pa\'is Vasco, Apartado 1072, 20080 San Sebasti\'an/Donostia, Spain}
\affiliation{Donostia International Physics Center (DIPC),
20018 San Sebasti\'an/Donostia, Spain}
\affiliation{Centro de F\'isica de Materiales CFM-Materials Physics Center MPC,
Centro Mixto CSIC-UPV/EHU, 20018 San Sebasti\'an/Donostia, Spain}

\begin{abstract}
Weyl semimetal is a three-dimensional material with a conical spectrum near an even number of point nodes, where two bands touch each other. 
Here we study spectral properties of surface electron states in such a system. We show that the density of surface states possesses a  
logarithmic singularity for the energy $\varepsilon \to 0$. It decreases linearly at the intermediate energy of surface electron states and  
approaches zero as $\sqrt{1-\varepsilon}$ for $\varepsilon \to 1$.    
This universal behavior is a hallmark of the topological order that offers a new wide range of applications. 
\end{abstract}
\maketitle
{\it Introduction}.-- New classes of matter known as topological insulators and Weyl semimetals are characterized
by the linear dispersion of low-energy electron excitations on the surface and in the bulk, respectively. The states on the surface of these so-called Dirac materials have a fixed spin orientation for each momentum. 
The electron states in topological insulators \cite{HK,QZ,M} are topologically ordered and 
protected by the time-reversal symmetry. A condition for the existence of Weyl semimetal is breaking of 
either inversion or time-reversal symmetry. The topological order manifests itself as massless Dirac modes 
propagating along the edge or the surface of topological insulators or in the bulk of Weyl semimetals. 
Study of the properties of surface electron states, being a hallmark of the topological order, 
enables one to clarify some features of the topological order in the bulk. It should be noted that the topological classification \cite{SRFL,SRFL1,K,V} of phase states in topological insulators has been 
extended to Weyl semimetals \cite{YN}. The transport features of Weyl semimetals \cite{ZB,PHXQ} related to the chiral anomaly as well as the spectrum of collective excitations \cite{LZ,PBP} were recently studied (see reviews \cite{HQ,TV}). Oscillations of the density of bulk states in 
Weyl semimetals in a strong magnetic field and their experimental signatures have been analyzed in Ref. \cite{AC}. 

In this paper, we examine the properties of Weyl semimetals when the time-reversal symmetry is preserved, 
while the spatial inversion symmetry is broken \cite{HB}, focusing on the density of {\it surface} states.  
This type of Weyl semimetal is studied in Ref. \cite{O} in which the spectrum of surface states is obtained. It has the form
\begin{equation}
\label{eq1}
E(k_{x},k_{y})=4t\sin\frac{k_{x}a}{4}\sin\frac{k_{y}a}{4}
\end{equation}
and is shown in Fig. 1. Here ${\bf k}=(k_{x}, k_{y})$ is the two-dimensional wave vector, $t$ is the integral of nearest-neighbor hopping, 
and $a$ is the lattice constant.
\begin{figure}
\includegraphics[width=1\columnwidth]{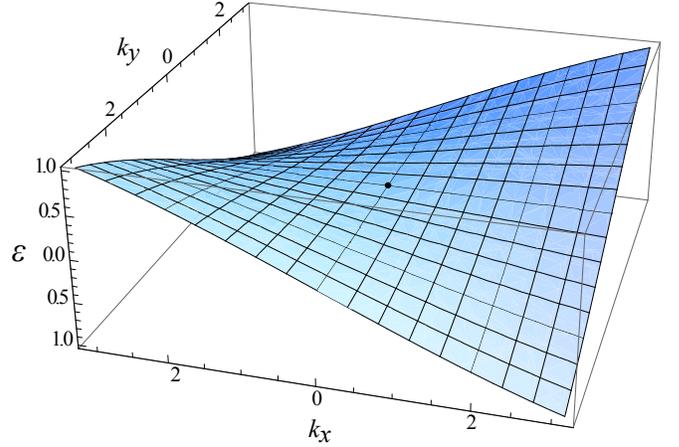}
 \caption{The spectrum of surface states in Weyl semimetal. The point denotes $k_{x}=k_{y}=0$.}
 \label{fig:Spectrum}
\end{figure}

{\it Density of surface states}.-- The density of surface states is defined by the integral over the surface 
Brillouin zone (BZ) $|k_{x}\pm k_{y}| \le \frac{2\pi}{a}$ of the delta function $\delta[E-E(k_{x},k_{y})]$ as follows  
\begin{equation}
\label{eq2}
N(E)=\int\limits_{BZ}\frac{d^{2}k}{(2\pi)^{2}}\delta[E-E(k_{x},k_{y})] .
\end{equation}
Having introduced the dimensionless quantities, i.e., the density of surface states $n(\varepsilon)=N(E)/N_{0}$ with $N_{0}=4/(\pi^{2}a^{2}t)$, 
the energy $\varepsilon=E/(2t)$, the wave vector components $(x,y)=(k_{x}a/4,k_{y}a/4)$, after the change of variables 
$\xi = x-y$, $\eta =x+y$, we obtain   
$$n(\varepsilon)=\frac{1}{4}\int\limits_{-\pi/2}^{\pi/2}d\xi \int\limits_{-\xi}^{\xi}d\eta \delta[\varepsilon - \cos\xi + \cos\eta ]=$$
\begin{equation}
\label{eq3}
=\int\limits_{0}^{1-\varepsilon}\frac{dt}{\sqrt{(1-t^{2})[1-(\varepsilon+t)^{2}]}}=\int\limits_{0}^{\varphi}\frac{d\alpha}{\sqrt{1-m\sin^{2}\alpha}}. 
\end{equation}
The amplitude $\varphi =\arcsin\sqrt{\frac{2(1-\varepsilon)}{2-\varepsilon}}$ and the parameter $m=1-(\varepsilon/2)^{2}$ of the elliptic integral of the first kind $F(\varphi,m)=\int\limits_{0}^{\varphi}\frac{d\alpha}{\sqrt{1-m\sin^{2}\alpha}}$
equal to the density of surface states depend on the energy $\varepsilon$ in the range
$0 < \varepsilon \le 1$. The asymptotic values of the elliptic integral of the first kind determine the behavior of the density of surface states for 
$\varepsilon \to 0$ and $\varepsilon \to 1$ as follows
\begin{equation}
\label{eq4}
n(\varepsilon)=
\begin{cases}
\ln\tan(\frac{\pi}{4}+\frac{\varphi}{2}), \,\,\,\,\,\,\,\,\,\,\,\,\,\, \,\,\, \, \varepsilon \to 0 \, \,\, \, (\varphi \to 
\frac{\pi}{2}),  \\
\sqrt{\frac{2(1-\varepsilon)}{2-\varepsilon}} \,\,\,\,\,\,\,\,\,\,\,\,\,\, \,\,\, \,\,\,\,\, \,\,\,
\,\,\,\,\,\,\,\,\,\varepsilon \to 1 \ .
\end{cases}
\end{equation}
The density of surface states versus the energy $\varepsilon$ in the range
$0.1 <  \varepsilon \le 1$ is shown in Fig. 2. 
It should be noted that the density of surface states in Eq. \ref{eq3} is valid in the energy range until the surface states merge with the bulk electron state continuum \cite{O}. This energy range is smaller than the interval $(0,1)$.  
\begin{figure}
\includegraphics[width=1\columnwidth]{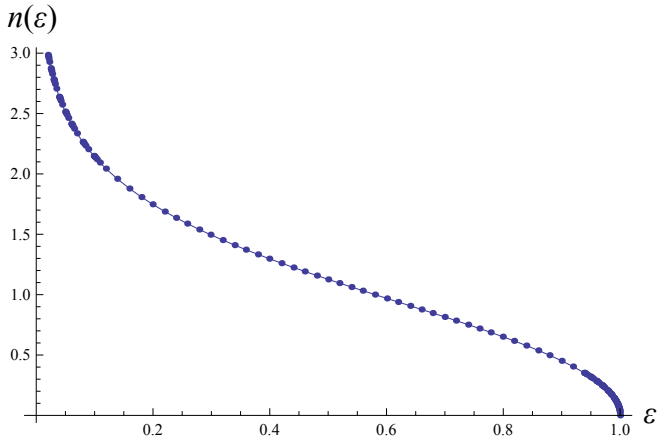}
  \caption{The density of surface states $n(\varepsilon)$ vs. the energy $\varepsilon$ in Weyl semimetal.}
  \label{fig:Density of states}
\end{figure}

{\it Discussions}.-- The density of states is determined by the dispersion of the electron spectrum and the spatial dimensionality $d$ of the considered problem. It is usually a non-decreasing function of energy. The 
exception is the one-dimensional case with the quadratic spectrum $E_{p}=p^{2}/2m_{e}$
when $N(E) \sim E^{-1/2}$. For comparison, the density of surface states in topological insulators is equal 
to $N(E)=\frac{E}{2\pi \hbar^{2}v_{F}^{2}}$, while it equals $N(E)=\frac{E^{2}}{2\pi^{2}\hbar^{3}v_{F}^{3}}$ for the bulk states in Weyl semimetals, where $v_{F}$ is the Fermi velocity. The restriction of the phase space affects the contribution to thermodynamic characteristics, 
e. g., by reducing the specific heat $C \sim T^{d}$ for Dirac materials, where $T$ is the temperature and $d=2,3$. In the considered two-dimensional problem, the origin of the decreasing function 
$n(\varepsilon)$ is the saddle point at the surface Brillouin zone center, which leads to the appearance of the Van Hove singularity for 
$\varepsilon =0$.     

Due to the decrease of the function $n(\varepsilon)$ in the intermediate region of energy, the density of surface states is generally more similar to the behavior in two one-dimensional perpendicular 
Dirac metals than to the behavior in a two-dimensional Dirac metal that exists on the surface of numerous 
topological insulators. In such a way, we confirm the approach \cite{TK,NCMT} based on the idea 
that one can consider this surface electron system as a composition of one-dimensional orthogonal Luttinger electron wires.  
We would also like to mention another example of very anisotropic quasi-one-dimensional Dirac electronic states 
on the surface of $Ru_{2}Sn{3}$ \cite{GEYZAFBBCB}. 

The knowledge of $N(\varepsilon)$ and the value $N(\varepsilon_{F})$ at the Fermi energy $\varepsilon_{F}$ is 
important for studying the internal electrostatic effects and the external gate-voltage effects \cite{BBBTMSH}. 
The quantum capacitance $C_{Q}$ per unit area of a two dimensional system is given, e. g., by $C_{Q}=e^{2}N(\varepsilon_{F})$, where $e$ is the electron charge. This could be used for experimental check of the features of the 
density of surface states in Weyl semimetals. The density of states also determines the dc conductivity $\sigma_{dc}$. 
Einstein's formula $\sigma_{dc}=e^{2}N(\varepsilon_{F})D$ expresses $\sigma_{dc}$ in terms of the density of states and the diffusion constant $D=v_{F}^{2}\tau$, where $\tau$ is the transport lifetime. 
Obviously, we have thereby a contribution of the surface conductivity to the total one in tanneling 
in Weyl semimetals. As for the distribution of spin degrees of freedom, 
bulk states for one Dirac node in Weyl semimetal resemble chiral quasi-spin configurations in graphene, 
while surface states in Weyl semimetal are analogs of helical distributions of spin directions in topological insulators. 
The energy spectrum and the spin texture of surface states can be experimentally studied using the tunneling spectroscopy technique, 
for which the behavior of the density of surface states is the key one. 
 
In conclusion, we have calculated the density of surface states in Weyl semimetals and have shown that it possesses the logarithmic singularity for $\varepsilon \to 0$ decreasing linearly at the intermediate energy of surface electron states and  
approaching zero as $\sqrt{1-\varepsilon}$ for $\varepsilon \to 1$. 
It resembles the behavior of the set of two orthogonal one-dimensional Dirac metals embedded in two-dimensional space.          

{\it Acknowledgement}.-- The authors are grateful to V. Ya. Demikhovskii, S. V. Eremeev, E. R. Kocharovskaya, and V. G. Tyuterev  
for useful discussions. This work was supported in part by RFBR Grant No. 14-02-00174 (V.A.V.) and by the University of the Basque Country UPV/EHU under Grant No. IT-756-13 (E. V. C.).

\end{document}